 \documentclass[pre,twocolumn,citeautoscript,amsmath,amssymb,superscriptaddress]{revtex4}

\usepackage[dvips]{graphicx}
\usepackage{dcolumn}
\usepackage{bm}
\usepackage{times}

\begin{document}

\title{Hydrogen storage in MOF-5: A van der Waals density functional theory
study}

\author{Yue Huang}
\affiliation{Key Laboratory of Advanced Microstructured Materials, MOE,
Department of Physics, Tongji University, 1239 Siping Road, Shanghai 200092, P.
R. of China}

\author{San-Huang Ke}
\email[Corresponding author, E-mail: ]{shke@tongji.edu.cn}
\affiliation{Key Laboratory of Advanced Microstructured Materials, MOE,
Department of Physics, Tongji University, 1239 Siping Road, Shanghai 200092,
P. R. of China}
\affiliation{Beijing Computational Science Research Center, 3 Heqing Road,
Beijing 100084, P. R. of China}


\begin{abstract}

Physisorption of hydrogen molecules in metal-organic frameworks (MOFs)
provides a promising way for hydrogen storage, in which the van der Waals (vdW)
interaction plays an important role but cannot described by the density functional
theory (DFT). By using the vdW density functional (vdW-DF) method, 
we investigate systematically the binding energies of hydrogen
molecules in MOF-5 crystal. We first examine the accuracy of this methodology by
comparing its results with those from the correlated quantum chemistry methods 
for several fragment models cut out from the crystal. Good comparable accuracy
is found. By performing calculations for the true crystal
structure adsorbing one or multiple H$_2$ in the primitive cell, 
we show that these fragment models which have been focused previously 
cannot represent well the property of the crystal which cannot, 
however, be dealt with by the quantum chemistry methods. 
It is found that the binding energy with the organic linker is much
smaller than with the metal oxide corner, which limits the H$_2$ loading. 
We show that this can be improved significantly (from 5.50 to 10.39 kJ/mol) by
replacing the H atoms of the organic linker with F atoms which cause extra
electrostatic interaction. 

\end{abstract}
\maketitle

\section{Introduction}

Hydrogen as a substitution for fossil fuel in applications of energy carrier has stimulated
extensive research because of its abundance on earth and especially its clean
outcome (water). However, a safe and efficient hydrogen storage technique still
remains as an obstacle in practice, in spite of various chemical and physical
schemes proposed \cite{Luo2004,Sandrock1999,Zacharia2007,Arean2003}, in which
hydrogen is stored in atomic and molecular forms, respectively.
In this regards, metal-organic frameworks (MOFs) was found to be a promising
materials family for hydrogen storage based on physical adsorption. So far, a series
of MOFs has been synthesized and their structural properties and hydrogen
loading under different temperatures and pressures have been investigated by
powder x-ray diffraction (XRD) and inelastic neutron scattering (INS) techniques
\cite{Rosi2003,Rowsell2005}. Compared with other schemes, 
MOFs have several outstanding advantages, such as very large surface/volume
ratio, tailorable chemical composition and structure, and the regular
high-symmetry periodic crystal structure.
MOF-5 as the first one in the series has become a particular focus of
MOFs study~\cite{Rosi2003, Rowsell2005, Sagara0412543, Civalleri2006, Kaye2007,
Cabria2008, Sillar2009, Maark2010, Hughes2011}.
It has been found experimentally that MOF-5 with the cubic
three-dimensional extended porous structure can achieve a gravimetric H$_2$
capacity up to 4.5\%wt at 78K \cite{Rosi2003, Rowsell2005}.
However, a common problem exists with MOFs for hydrogen storage: Around room 
temperature the H$_2$ binding energy is still not large enough.
How to increase the adsorption
enthalpy to above 15 kJ/mol \cite{Bhatia2006,van-den-Berg2008,Lochan2006}
under room temperature has become a challenge in practice, which may be possible
by tailoring the chemical species and atomic structure. To
achieve this goal we need to first gain a full understanding of H$_2$
interaction with MOF-5 on the different adsorption sites in the internal space of MOF-5.

To accurately calculate these interactions is, however, a big challenge to theoretical studies
because the van der Waals (vdW) interaction plays an important role in the physisorption, which is a quantum dispersion force and is
related to the nonlocal correlation effect. This nonlocal correlation is absent
in the density functional theory (DFT) and can only be described by the
many-body perturbation theory or the correlated wavefunction methods. The lowest level of theory which can deal with
the vdW interaction is the M{\o}ller-Plesset second-order perturbation (MP2)
theory \cite{Head-Gordon1988}. More sophisticated methods include the coupled-cluster method,
such as the coupled-cluster singles and doubles and noniterative triples
(CCSD(T)) \cite{Raghavachari1989479}. These correlated quantum chemistry methods
also face challenges for the study of hydrogen storage in MOFs. First, these
methods cannot deal with periodic crystal structures. Consequently, one has to use some
fragment models cut out from the crystal structure to mimic the different parts
of MOFs \cite{Mueller0517974, Lochan2006, Buda0610479}. Considering the long-range nature of
the electrostatic and vdW forces present in the physisorption,
whether and to what extent these fragment models can represent the behavior of
the crystal is a question to answer. These correlated methods are also
computationally very demanding and cannot deal with large fragment models.
Furthermore, as usually implemented with the Gaussian basis functions, these
methods have a strong basis set dependence and, therefore, the result becomes
reliable only when a very large basis set is used or some elaborate correction
techniques are applied \cite{Thonhauser07125112}.

To simply add empirically a $1/r^{6}$ term to conventional DFT calculations,
as did in the DFT+Dispersion (DFT+D) method \cite{Grimme2004,Grimme061787},
may be a solution for studying the H$_2$ adsorption in the full crystal structure, but its accuracy and
reliability of result still needs to be examined
\cite{Sillar2009,Mueller0517974}.
On the other hand, in the vdW-DF method developed by Dion {\it et al.}
\cite{Dion04246401,Cooper2010prb}
the vdW interaction can be included perturbatively in a seamless fashion into an
first-principles Kohn-Sham DFT calculation. This is achieved by expanding the
correlation energy to the second order in terms of a carefully chosen quantity
for the long-range part of the correlation functional.
This method has been applied to different separational systems
\cite{Dion04246401,Thonhauser07125112} and shown reasonable accuracy.
For the extended crystal structure of MOFs interacting with H$_2$, whether and
to what extent the binding energy can be determined by the vdW-DF method
has not been studied in the literature, to our best knowledge.

In this work, we adopt the vdW-DF method to study systematically the binding energy of
H$_2$ in MOF-5 crystal for two major adsorption sites. 
We first test the accuracy of the methodology by considering
the fragment models already studied in the literature. The comparison with
the MP2, CCSD(T), and DFT+D calculations shows a good accuracy which is better
than the empirical DFT+D method. Calculations for the full crystal
structure, however, show that these fragment models cannot represent well the
hydrogen-storage property of the crystal due to the long-range interaction of
H$_2$ with the environment. Besides the single H$_2$ binding behavior, 
multiple H$_2$ adsorption is also considered to investigate the effect of H$_2$
density. Furthermore, to illustrate the possibility of enhancing the
average H$_2$ binding energy in MOF-5 crystal, we consider the substitution of F
atoms for the H atoms of the organic linker part.
It is found that in this way the H$_2$ binding energy can be almost doubled due to the
extra electrostatic interaction, indicating a possible way to improve the
performance of MOFs for hydrogen storage. 

\begin{figure}[tb]
(a)\includegraphics[width=8.0cm,clip]{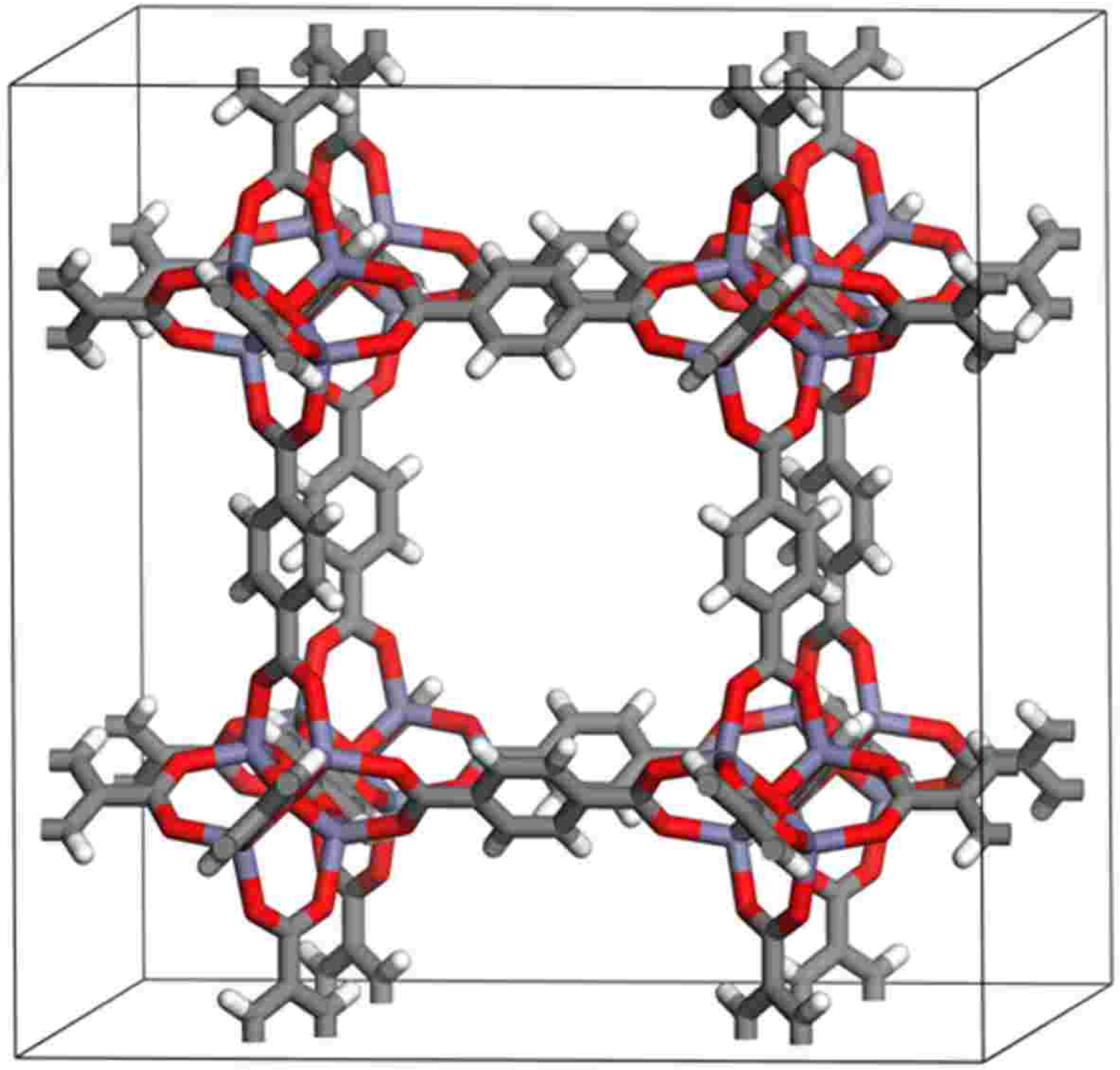} \\
(b) \includegraphics[width=8.0cm,clip]{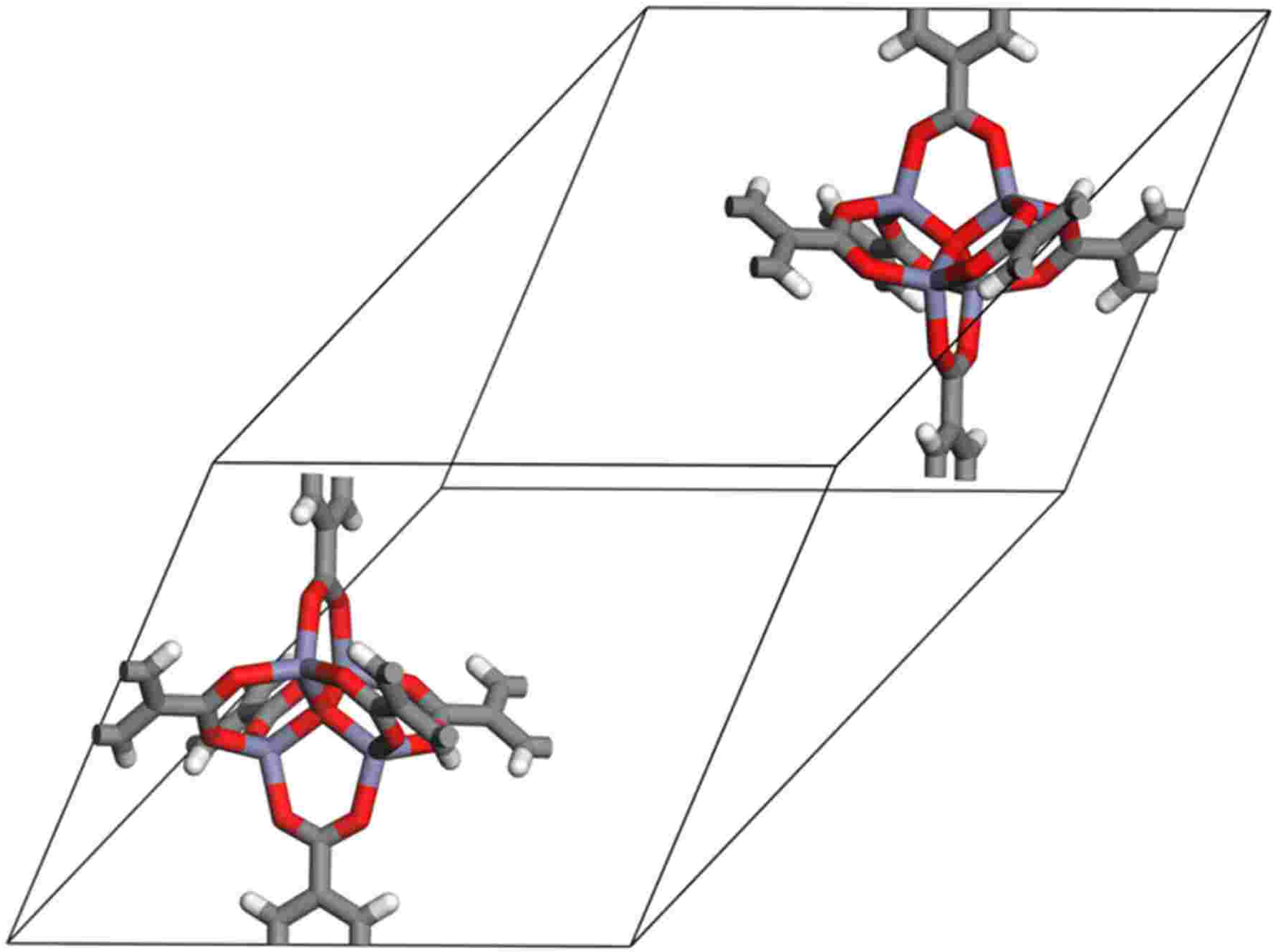}
\caption{
\label{fig:1}
(a) The conventional cell (424 atoms) and (b) primitive cell (106 atoms) of the MOF-5 crystall
which has the symmetry of Fm$\bar{3}$m and contains
eight and two formula units (OZn$_4$O(BDC)$_3$), respectively.
Red, blue, gray, and white denotes O, Zn, C, and H atoms, respectively. }
\end{figure}

\section{Theory, models, and computational details}

Electrostatic force and vdW force are the sources of the physisorption for the
hydrogen storage in MOFs. Therefore, both of them are important for getting
computationally accurate binding energy of H$_2$. Since the vdW interaction is absent in DFT
we adopt the recently developed vdW-DF method which includes the nonlocal
electronic correlation effect into the Kohn-Sham DFT calculation in a seamless
fashion\cite{Dion04246401}. This is achieved by expanding the correlation energy
to the second order in terms of a carefully chosen quantity for the long-range part of the correlation
functional. In this way, long-range dispersion effects are included as a perturbation to the
correlation term of the local or semilocal density approximation \cite{Cooper2010prb}.
This method has been successful in describing various groups of materials,
ranging from molecules to surfaces and bulk materials
\cite{Dion04246401,Thonhauser07125112,Cooper2010prb}.
Considering the limitation of the sophisticated quantum chemistry methods for
periodic systems, this method may provide a useful tool for the study of
hydrogen storage with MOF crystals, which has not been reported in the literature
to our best knowledge.

\begin{figure}[tb]
\includegraphics[width=8.0cm,clip]{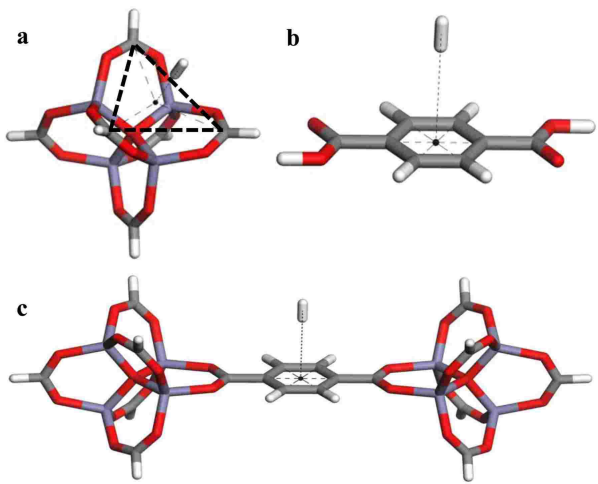}
\caption{
\label{fig:3}
Optimized structures of the three fragment models: (a) the corner model with a
H$_2$ adsorbed on the cup site, i.e., above the C$_3$ plane indicated by the
dashed-line triangle, (b) the BDC model with a H$_2$ adsorbed above the benzene
ring, and (c) the edge model which is made up of the two corner models linked by one BDC linker.
}
\end{figure}

The crystal structure of MOF-5 is composed of inorganic [OZn$_4$]$^{6+}$ octahedral groups linked
by 1,4-benzenedicarboxylate (BDC) organic linkers, forming a highly microporous cubic framework,
as shown in Fig.~\ref{fig:1}(a). It has a symmetry of face-centered cubic Fm$\bar{3}$m with
an experimental lattice constant of 25.89 {\AA} \cite{Rosi2003} for the
conventional cell. The conventional cell contains eight formula
units Zn$_4$O(BDC)$_3$ (424 atoms in total) while the primitive cell contains
only two (106 atoms in total), as shown in Fig.~\ref{fig:1}.
The large empty hole formed can be an ideal space for hydrogen storage where
the physisorption takes place. Specifically, 
both the OZn$_4$ cluster in the corner and the organic linker in the pillar can interact with H$_2$
molecules with different interaction strengthes. Here, we consider two 
adsorption sites, one is above the C$_3$ plane of the corner (the cup site, see
Fig.~\ref{fig:3}(a)) and the other is above the benzene ring of the organic
linker (linker site, see Fig.~\ref{fig:3}(b)). Previous theoretical and
experimental studies showed that the cup site is the most favored one and will be taken up 
by hydrogen molecules first, then the linker site will be occupied.
\cite{Yildirim05215504, Rowsell2005}

In order to examine the reasonability of fragment models 
in representing the different parts of the full crystal,
three fragment models, corner model, BDC model, and edge model, are
considered, which are used to represent the metal oxide corner, organic
linker, and one edge of MOF-5, respectively (see Fig.~\ref{fig:3}).
These models are also used to examine the accuracy of the vdW-DF method by
comparing its results to the existing results from the correlated quantum
chemistry methods as well as DFT+D. 

Our calculations are carried out by using the pseudopotential plane-wave
formulism of the vdW-DF theory, as implemented in the Quantum ESPRESSO package\cite{QE-2009}.
The O, Zn, C, and H atoms are described by the ultrasoft pseudopotentials
\cite{Vanderbilt1990,Laasonen1993} and a kinetic energy cutoff of 30 Ry is used for the wavefunction
expansion. Considering the fact that the force acting on H$_2$ is much smaller than
those in usual bonded systems, we set a very high total-energy convergence
criteria, $1.0\times10^{-7}$ Ry, for the selfconsistent vdW-DF calculation to obtain accurate forces on
atoms. The structures of the fragment models (with and without H$_2$) are optimized by minimizing
the forces on each atom to be smaller than $1.0\times10^{-4}$ Ry/Bohr. The structures
of the crystall (with and without H$_2$) are optimized by performing full
variable-cell relaxations until the forces on atoms are smaller than
$1.0\times10^{-4}$ Ry/Bohr and the change in total energy of the primitive cell
between two neighboring variable-cell steps is smaller than $1.0\times10^{-5}$ Ry.
A 4$\times$4$\times$4 Monkhorst-Pack k-point mesh is used to sample the
Brillouin zone of the crystall.

\begin{figure}[tb]
(a) \includegraphics[width=8.0cm,clip]{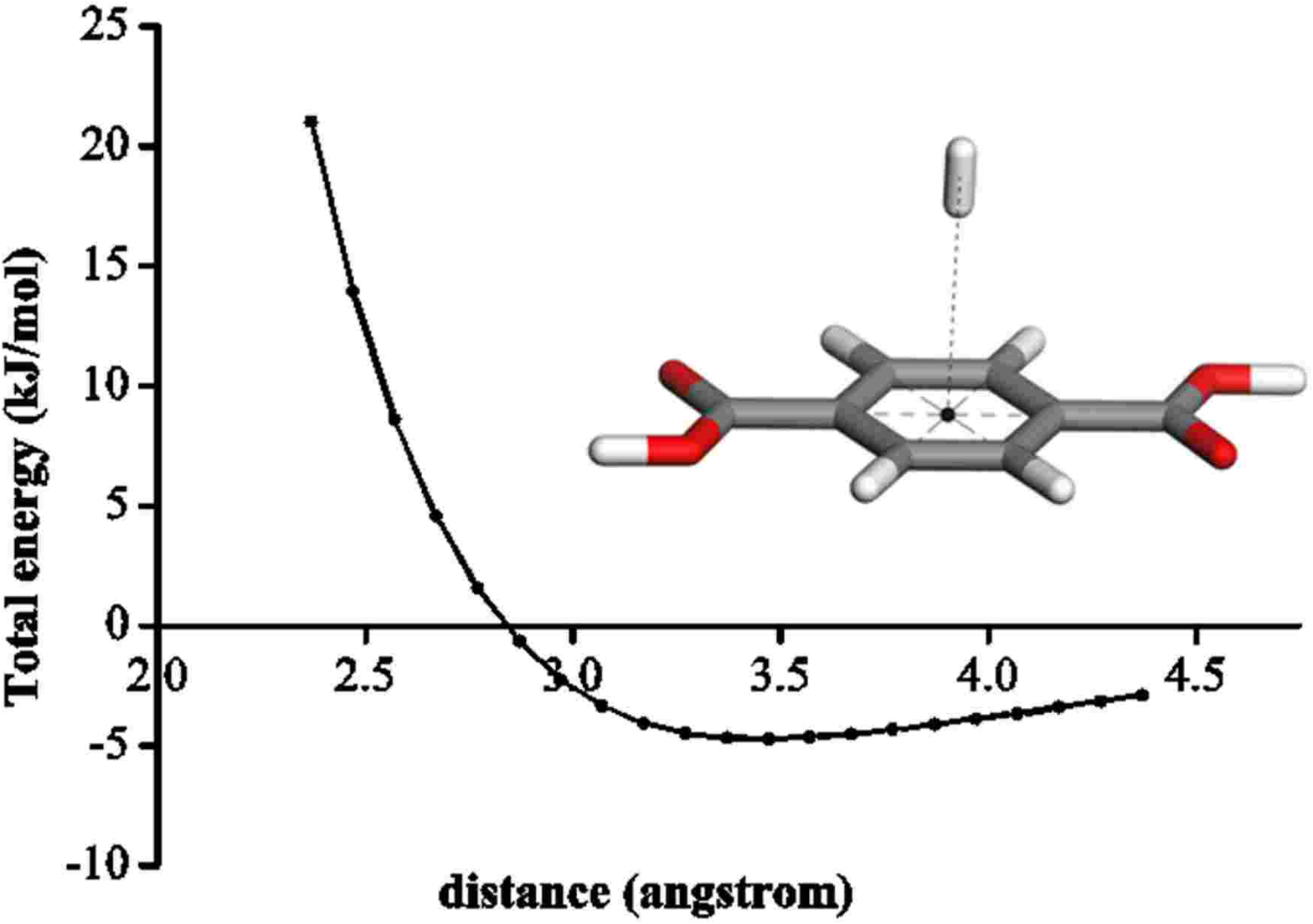} \\
(b) \includegraphics[width=8.0cm,clip]{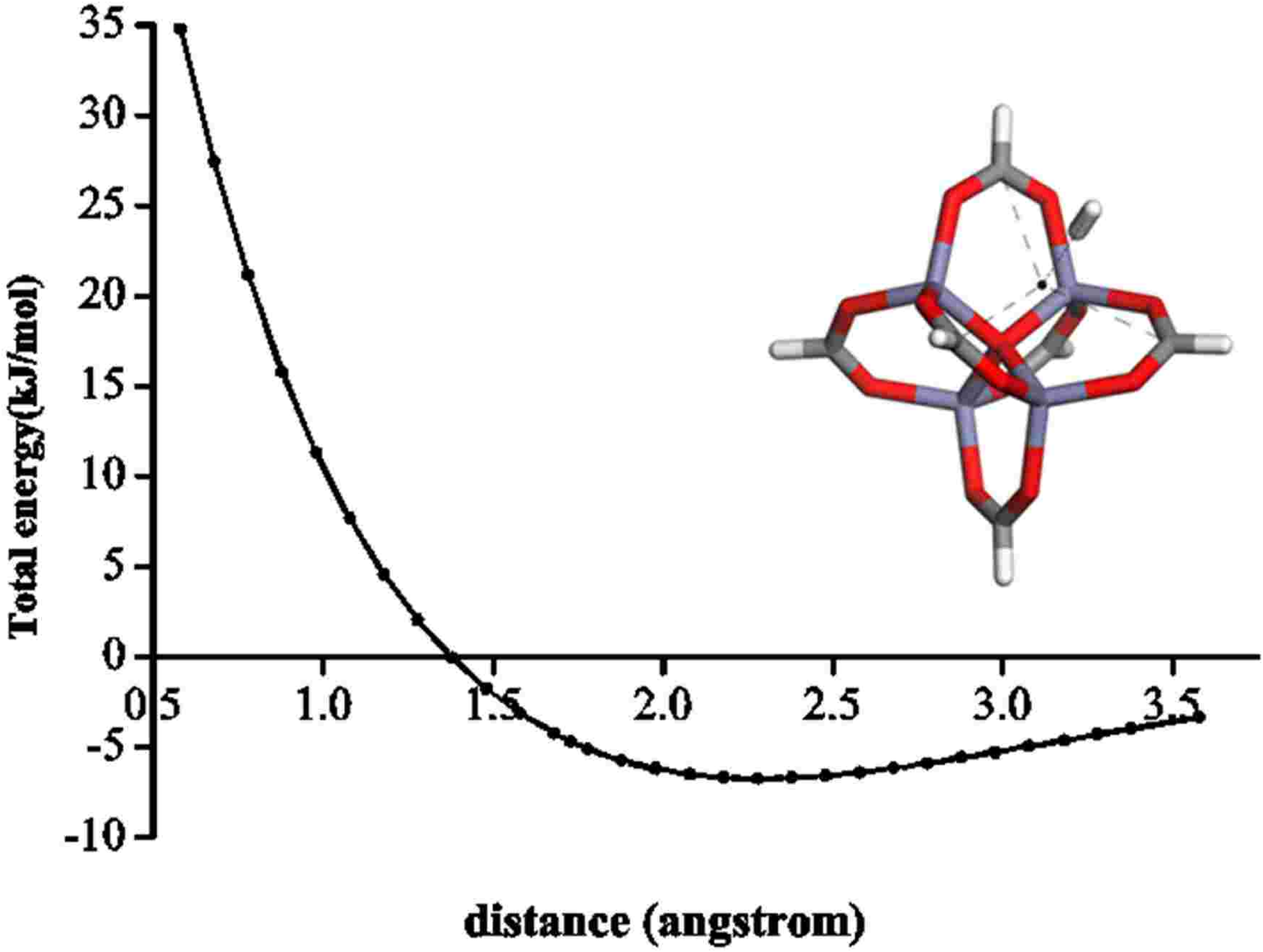}
\caption{\label{fig:4}
(a) The total energy as a function of the adsorption distance between the center of the
unrelaxed H$_2$ and the C$_6$ plane of the unrelaxed BDC model.  
The H$_2$ and BDC model are fully optimized before the calculation.
(b) Similar plotting for the corner model with a H$_2$ adsorbed on the
cup site. The adsorption distance is between the center of the H$_2$ and the C$_3$ plane.
}
\end{figure}

\section{Results and Discussion}

\subsection{Structure optimization with a single H$_2$ adsorbed}

We first consider a single H$_2$ molecule adsorbed on the fragment models
and in the MOF-5 crystal structure. 
Since the physisorption is quite weak, the most stable
adsorption configuration needs to be determined very carefully. For the fragment
models we do the structure optimization in two steps. First, we perform a series
of single point total-energy calculations for different adsorption distances
within a roughly chosen range (see Fig.~\ref{fig:4}). Second, based on the
lowest-energy adsorption distance obtained we then do a full structure relaxation 
with a very high convergence criteria, 
as mentioned previously, to ensure an accurate final adsorption configuration. The
final optimized adsorption distances are also used as the initial input for the
calculation of the crystal structure.
For example, for the BDC model the result from the first step (see 
Fig.~\ref{fig:4}(a)) gives a lowest-energy distance of 3.40{\AA} in line with the typical vdW
interaction distance \cite{Lochan2006}. The final fully optimized distance from
the second step is 3.48\AA. For the corner model, the total-energy curve in
Fig.~\ref{fig:4}(b) shows a lowest-energy distance of 2.28{\AA} and the fully
relaxed one is 2.01\AA. The corresponding results for the crystal
structure are 3.49{\AA} and 2.06{\AA} for the two adsorption sites, respectively.
The full results are listed in Table~\ref{Table:1}. 

\begin{table}[h] 
\centering
\begin{tabular}{c c c c c c}
\hline
\hline
          & corner+H$_2$ & BDC+H$_2$ & edge+H$_2$ & p(c)+H$_2$ &p(l)+H$_2$ \\
\hline
distance  &  2.01        & 3.48      & 3.46       & 2.06           & 3.49  \\
vdW-DF    &  8.68        & 4.55      & 5.31       & 9.39           & 5.50  \\
CCSD(T)   &  --          & 4.32$^b$  &  --        & --             & --    \\
MP2       &  7.50$^a$    & 5.41$^b$  & 5.10$^a$   & --             & --    \\
PBE+D     &  5.30$^a$    & --        &   --       & 9.60$^a$       & --    \\
\hline
\hline
\end{tabular}
\caption{ The calculated adsorption distance (distance, in \AA) and binding energy (vdW-DF, in kJ/mol) of
a single H$_2$ molecule adsorbed on the cup site or linker site of the fragment
models (corner+H$_2$, BDC+H$_2$, and edge+H$_2$) and crystal structure
(p(c)+H$_2$, p(l)+H$_2$). Results of binding energy from some previous
calculations (including CCSD(T), MP2, and PBE+D) are also listed for a
comparison.\\ 
$^a$  data from Reference \cite{Sillar2009},\\
$^b$  data from Reference \cite{Sagara05014701}. 
}
\label{Table:1}
\end{table}

\subsection{Binding energy of a single H$_2$ and the accuracy of the vdW-DF method}

We first examine the accuracy of the vdW-DF method by performing calculations
for the fragment models and compare the results of H$_2$ binding energy to the
existing results from the correlated quantum chemistry methods.
The calculated results are listed in Table.~\ref{Table:1} together with reliable results from CCSD(T), MP2,
and DFT+D using PBE functional (PBE+D). Sagara {\it et al} has reported H$_2$ binding energy for the BDC model using
the CCSD(T) and MP2 methods adopting the large QZVPP
basis set \cite{Raghavachari1989479} and got results of 4.32 and 5.41
kJ/mol, respectively. Our vdW-DF result is 4.55 kJ/mol which is very close to
the CCSD(T) one. For the corner model, a MP2 calculation \cite{Sillar2009} using a basis set of
aug-cc-pVQZ plus diffuse functions gave a result of 7.50 kJ/mol while a PBE+D
calculation \cite{Sillar2009} adopting a plane-wave basis set with 400 eV kinetic-energy cutoff
showed 5.30 kJ/mol. Our result of 8.68 kJ/mol is
closer to the MP2 result than the PBE+D one does. For the edge model, the present
result of 5.31 kJ/mol is also quite close to the MP2 result 5.1 kJ/mol obtained
in Ref.~\cite{Sillar2009}.
From the above comparison, we can conclude that the vdW-DF method can describe
well the physisorption of H$_2$ molecule in MOF-5 and its accuracy is
comparable to that of the MP2 and CCSD(T) methods and seems better
than the DFT+D. An outstanding advantage of the DFT-based methods is that,
unlike the correlated quantum chemistry methods, they
are not so strongly dependent on the basis set used. For the latter, although the theories themselves are
more accurate but their results are very scattered in the literature,
varying largely with different basis sets used, and only
become reliable when the very big basis sets (as mentioned above) are adopted.
Consequently, even for the small fragment modes they are computationaly very demanding.

\begin{figure*}[tb]
(a)\includegraphics[width=7cm,clip]{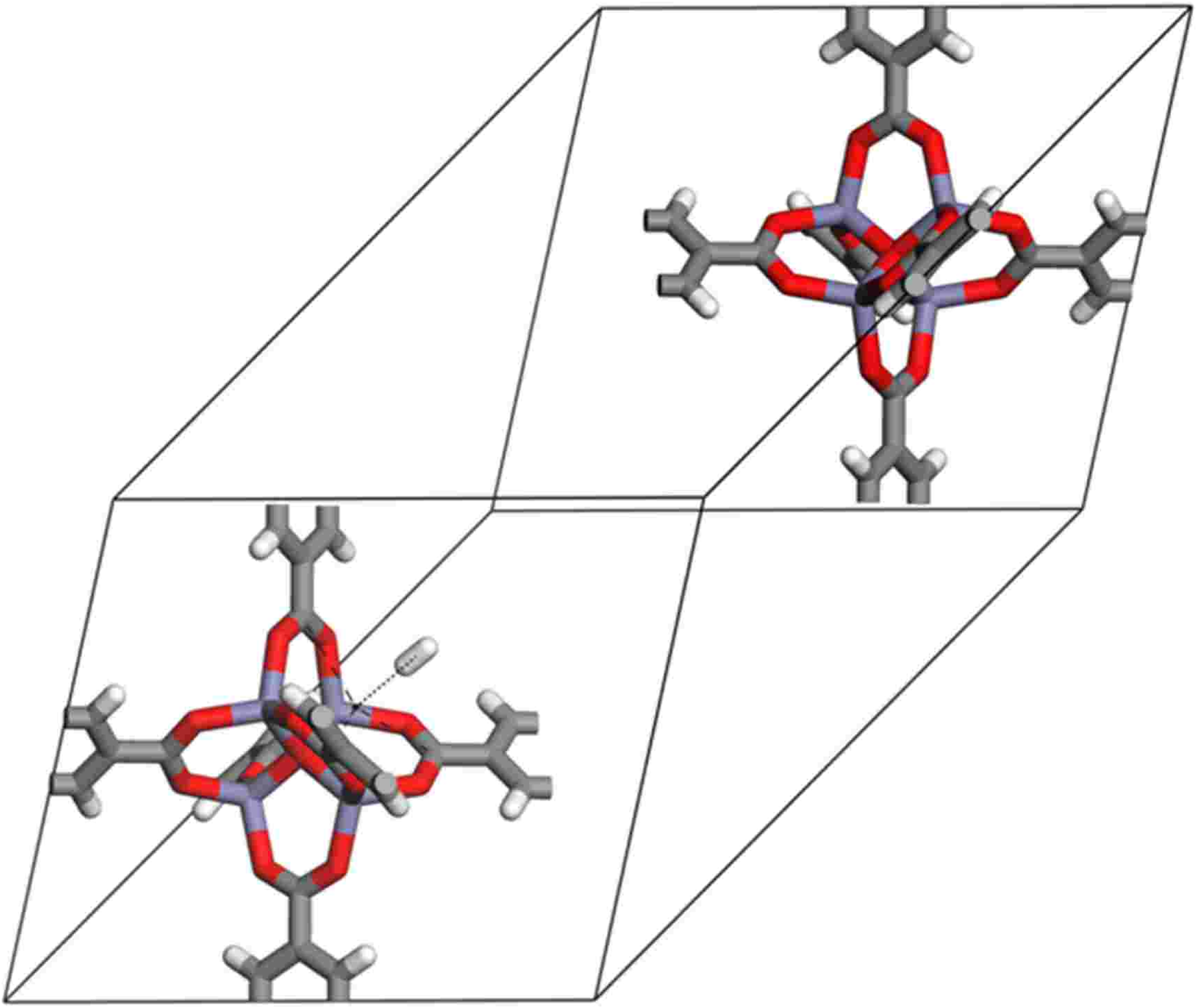}
(b)\includegraphics[width=7cm,clip]{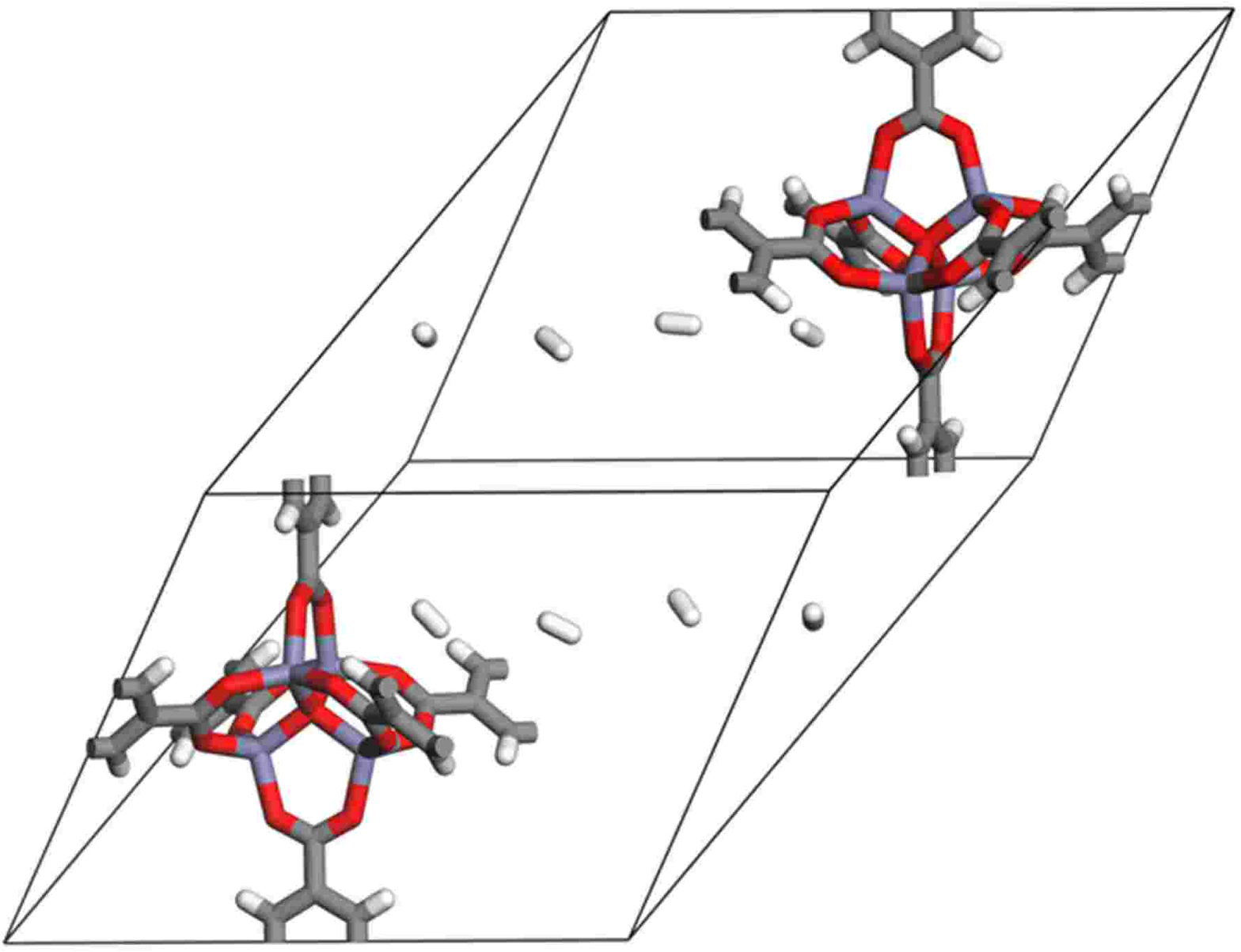} \\
(c)\includegraphics[width=7cm,clip]{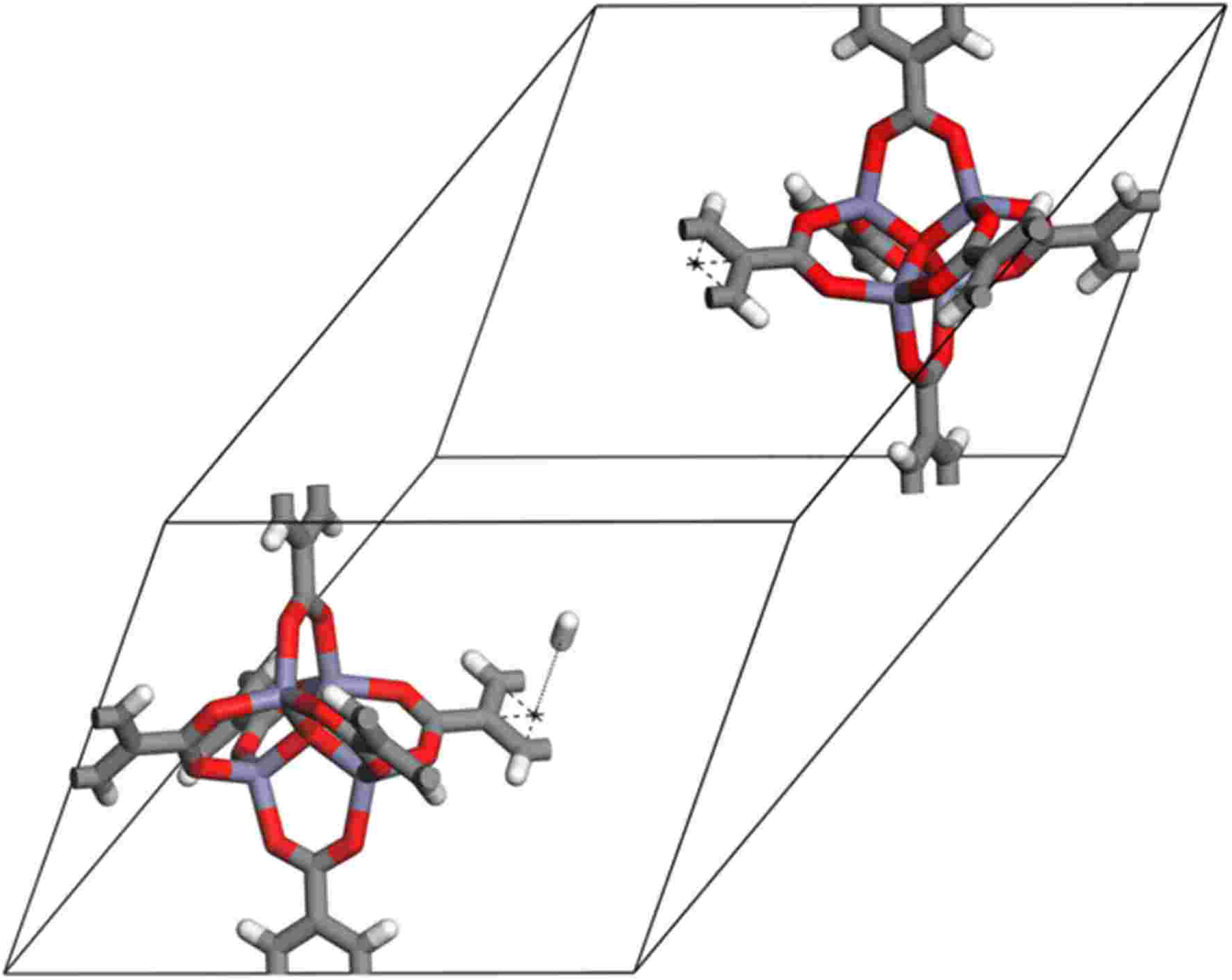}
(d)\includegraphics[width=7cm,clip]{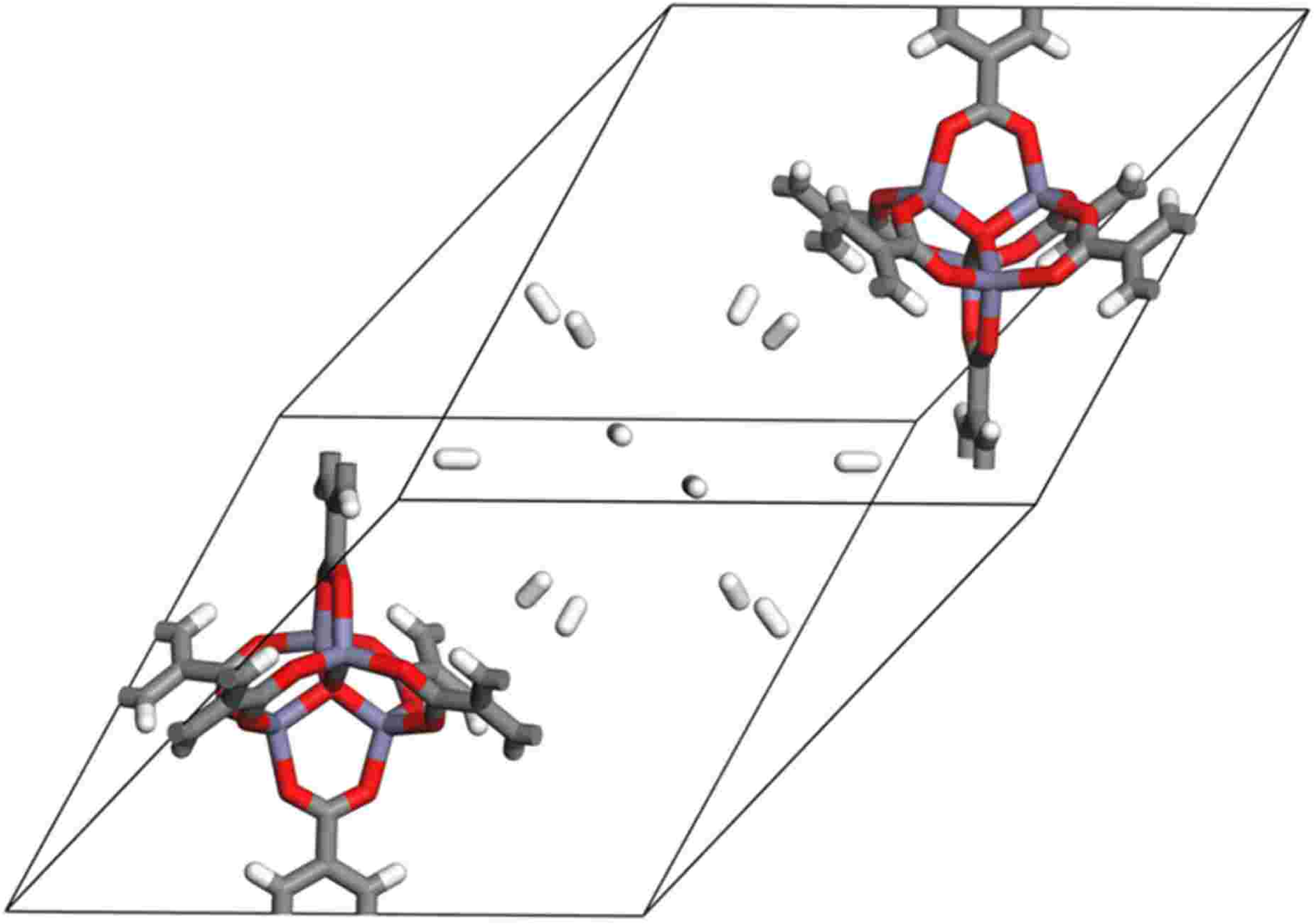}
\caption{\label{fig:8} The MOF-5 primitive cell with (a) one H$_2$ adsorbed on
the cup site, (b) 8 H$_2$ adsorbed on the cup sites, (c) one H$_2$ adsorbed
on the linker site, and (d) 12 H$_2$ adsorbed on the linker sites. 
}
\end{figure*}

%

\subsection{Fragment models vs. crystal structure}

As one can see in Table~\ref{Table:1}, the fragment modes and the crystal
structure give quite different results of binding energy for a single H$_2$
adsorbed.
The binding energy with the corner model (corner+H$_2$) is 0.71 kJ/mol lower than that with the
crystal structure (p(c)+H$_2$) and the binding energy with
the BDC model (BDC+H$_2$) is 0.95 kJ/mol smaller than the crystal one (p(l)+H$_2$).
This difference is due to the effect from the extended environment in the crystal
structure which may provide an accumulated electrostatic and vdW forces because
of their long-range nature. Additionally, there will be some charge
transfer between a fragment and its environment, which will also change the
strength of the physisorption. Our result shows that this effect is particular
significant when the H$_2$ molecule is adsorbed on the organic linker part:
Without connecting to the metal oxide corner part the BDC model underestimates
the binding energy by about 20\%. After this connection is made, as in the case
of the edge model (edge+H$_2$), then the result becomes closer to (but still
smaller than) that of p(l)+H$_2$.
However, this connection will increase significantly the size of the model and make the
higher-order quantum chemistry CCSD(T) calculation a big challenge.

Our calculation demonstrates that, compared to the true crystall structure
calculation,  the small fragment modes can still capture the main
feature of the physisorption of H$_2$ on the different adsorption sites: Both
show that the binding on the cup site of the OZn$_4$ corner is much stronger than that on the linker
site. This has also been comfirmed experimentally by
inelastic neutron scattering spectroscopy \cite{Murray091294} and neutron powder
diffraction measurement \cite{Yildirim05215504}. The physics underlying is that the
metal ion induced charge polarization around the corner causes stronger electrostatic force on the
H$_2$. However, our calculation shows that the small fragment models ignoring the influence of the adjacent components are not quantitatively
sufficient to represent the different parts of MOFs for investigating the
hydrogen storage with MOFs. To get accurate H$_2$ binding energies for the different
adsorption sites, calculations with the true periodic boundary conditions
for the crystal structure are needed. Besides the underestimated binding energy
for a single H$_2$ adsorbed, the fragment models also give a different trend for
multiple H$_2$ adsorbed, as will be discussed later.

\begin{table*}[t]
\centering
\begin{tabular}{c c c c c c c}
\hline
\hline
               & corner+H$_2$ &     & corner+4H$_2$ & BDC+H$_2$  &       & BDC+2H$_2$ \\
\hline
distance       &  2.05        &     & 1.93          & 3.48       &       & 3.46       \\
vdW-DF         &  8.68        &     & 8.78          & 4.55       &       & 4.56       \\
\hline
\hline
               & p(c)+H$_2$  & F-p(c)+H$_2$  & p(c)+8H$_2$  & p(l)+H$_2$ & F-p(l)+H$_2$  & p(l)+12H$_2$ \\
\hline
distance       &   2.06      & 2.05          &  2.07        &  3.49      & 3.45 &  3.48  \\
lattice const. &  18.81      & 18.73         &  18.79       &  18.81     & 18.72&  18.79 \\
vdW-DF         &   9.39      & 9.82          &  9.05        &  5.50      & 10.39&  5.14  \\
\hline
\hline
\end{tabular}
\caption
{Calculated average binding energy (vdW-DF, in kJ/mol) per H$_2$ for one and multiple H$_2$
adsorbed on the fragment models (the first part of the table) and in the
crystall structure (the second part of the table). The corresponding adsorption distance and
the lattice constant are also listed. 
F-p(c)+H$_2$ (F-p(l)+H$_2$) denotes one H$_2$ adsorbed on the cup (linker) site
in the crystall structure with the fluorine substitution, as described in the
text.
}
\label{Table:2}
\end{table*}

\begin{figure*}[tb]
(a)\includegraphics[width=7.0cm,clip]{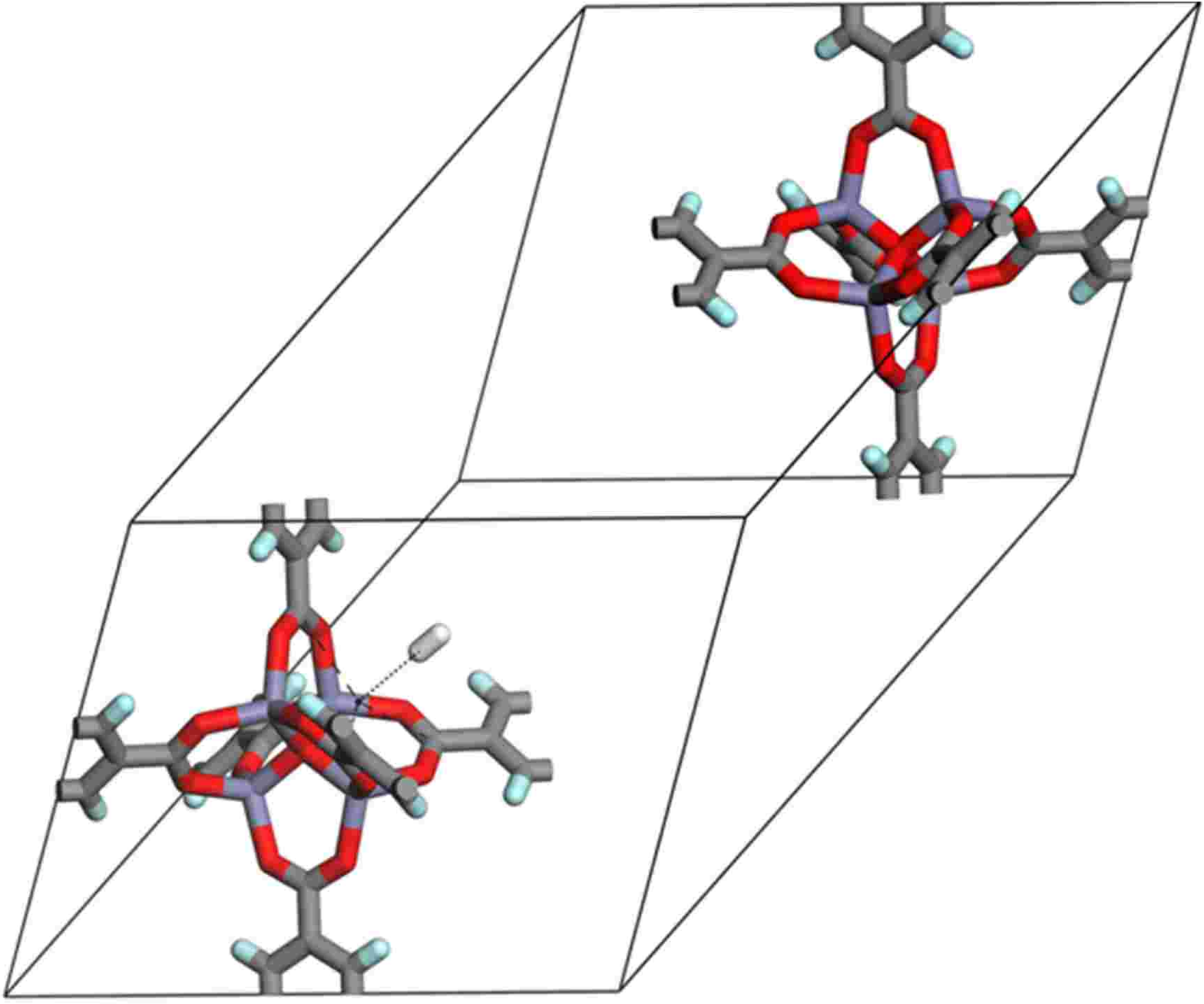}
(b)\includegraphics[width=7.0cm,clip]{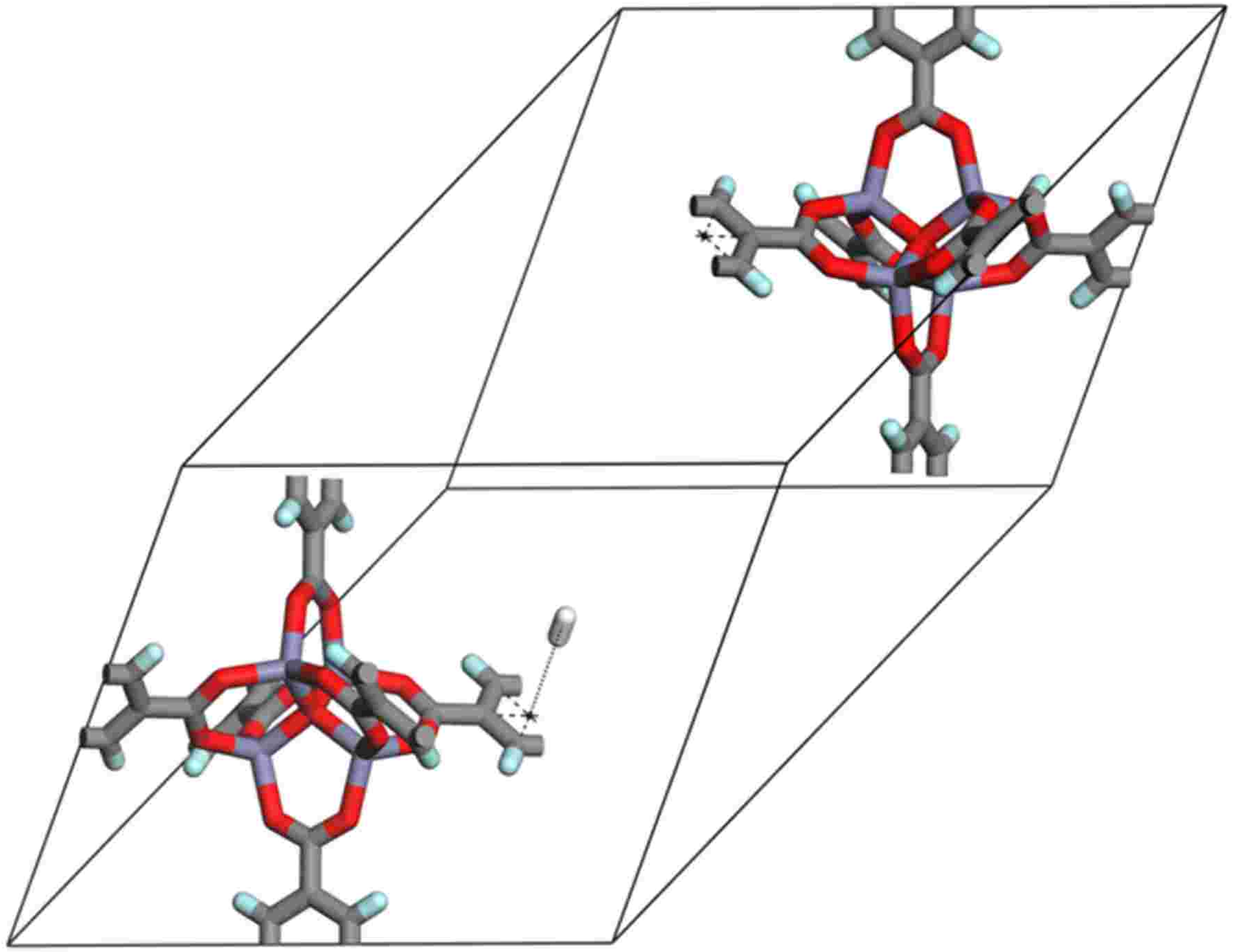}
\caption{\label{fig:9}
Primitive cell with the H atoms of the organic linkers replaced by F atoms: (a) with
one H$_2$ adsorbed on the cup site, (b) with one H$_2$ absorbed on the linker
site.
}
\end{figure*}

\subsection{Effect of multiple H$_2$ adsorbed }

In the practice of hydrogen storage many H$_2$ molecules will be adsorbed in
the primitive cell of a MOF crystall. Here, we investigate the effects of hydrogen density on the
binding energy by considering multiple H$_2$ molecules adsorbed in the MOF-5
crystall, on the corner site (p(c)) and the linker site (p(l)), as shown in Fig.~\ref{fig:8}.
The lattice constant and the average binding energy per H$_2$ for difference H$_2$ densities are listed in Table~\ref{Table:2}. 
One can see that when the number of H$_2$ adsorbed in the primitive cell increases,
the lattice constant actually decreases a little bit (by 0.02 \AA) and the binding
energy also decreases by 0.34 kJ/mol and 0.36 kJ/mol for the p(c) and p(l), respectively.
This small decrease of the lattice constant implies that there exists an attractive 
force among the adsorbed H$_2$ molecules, which increases with the increaseing
H$_2$ density. This behavior is different from the case of chemisorption where
an increase in adsorbate density will usually lead to an expansion of the
volume.
For the highest H$_2$ density considered in the present work, the smallest
H$_2$...H$_2$ distance is around 4.1{\AA} which is quite close to that in
the bulk H$_2$ solid (3.8 \AA) \cite{silvera1980}. For even higher H$_2$ density
comparable to that in the solid state, a previous calculation by Yildirim {\it et al} \cite{Yildirim05215504} 
showed that a unique three-dimensional interlinked H$_2$ nanocage will be
formed, which will enhance further the interaction among the H$_2$ molecules. 
Since the attraction among the H$_2$ molecules is in the opposite direction to
their binding to the MOF-5 crystall, it reduces the binding energy consequently.

As a comparison to the crystall case, we also check the multiple H$_2$ effect using the fragment models.
As shown in Table~\ref{Table:2}, for the corner model, when the four cup sites are occupied
by H$_2$, the average binding energy actually increases by 0.1 kJ/mol compared to the case of only one H$_2$
adsorbed. For the BDC model, this change is much smaller but is still increased
(by 0.01 kJ/mol) with H$_2$ adaorbed on both sides of the benzene ring compared
to only on one side.
This qualitative difference between the crystall and the fragment
models for multiple H$_2$ adsorption can be understood by considering the fact
that in the fragment models the H$_2$ molecules are adsorbed only on the surface of
the cluster, and therefore, the interlink between H$_2$ molecules adsorbed in a
cage of the crystall structure is absent. In this case, when multiple H$_2$
molecules are adsorbed the synergistic effect of electrostatic and vdW
forces leads to the small increase of the binding energy.
Thus, our calculation shows that, compared to the true crystall stucture, the fragment models not only underestimates the
binding energy when a single H$_2$ molecule is adsorbed but also give an
opposite trend of the binding energy when multiple H$_2$ molecules are adsorbed.

Another important effect of multiple H$_2$ adsorption is an perpendicular-to-parallel rotation of the
H$_2$ molecules adsorbed on the cup site: When only one H$_2$ is
adsorbed, it prefers to be perpendicular to the C$_3$ plane (see Fig.~\ref{fig:8}(a)). However,
when multiple H$_2$ molecules are adsorbed on the cup site, their adsorption
configuration will undergo an perpendicular-to-parallel transition and becomes parallel to
the C$_3$ plane (see Fig.~\ref{fig:8}(b) ). This transition reflects the interaction between 
the H$_2$ molecules adsorbed on the same corner unit.

\subsection{Substitution of F atoms for the H atoms of the organic linker}

As is already shown, the H$_2$ binding energy with the organic linker is  
significantly smaller than with the metal oxide corner, which largely limits the H$_2$
loading for hydrogen storage application. 
The reason is that the electrostatic force coming
from the OZn$_4$ cluster is much stronger than the vdW force coming from the organic linker. 
This implies a way for improving the average binding energy of H$_2$ molecules
in the MOF-5 crystall: It may be achieved by creating extra polarization in the
organic linker part and therefore enhancing the binding energy by the induced
electrostatic force. 

Here we consider a possibility that the 4 hydrogen atoms of an organic linker are
replaced with 4 fluorine atoms (see Fig.~\ref{fig:9}) which have a much stronger
electronegativity leading to a larger charge transfer and, therefore, will give rise to
extra polarization. Fluorine atom has also a small mass for a good gravimetric hydrogen
loading. 
%
%
The results of binding energy after the fluorine substitution are also given in
Table.~\ref{Table:2} (F-p(c)+H$_2$ and F-p(l)+H$_2$).
One can see that for one H$_2$ adsorbed on the cup site the binding
energy increases slightly from 9.39 kJ/mol to 9.82 kJ/mol. However, for one H$_2$ adsorbed on
the linker site the binding energy is enhanced significantly from 5.50
kJ/mol to 10.39 kJ/mol which is now even larger than the binding energy with the
corner, showing that the proposed extra electrostatic interaction is really
working. 
This result casts a light on possible ways for improving the performance of MOFs
for hydrogen storage.

\section{Summary}

By using the vdW-DF method implemented with plane-wave basis functions
we have investigated systematically the physisorption of hydrogen molecules in
MOF-5 crystall as used for hydrogen storage.
We first examed the accuracy of this methodology by comparing the results 
for three framgent models to those from the correlated quantum chemistry
methods, CCSD(T) and MP2, as well as from the DFT+D method.
Good accuracy was found, which is comparable to that of the CCSD(T) and MP2 with
very large basis sets. By comparing the results for the fragment modes with those for the
true crystall structure we have shown that the fragment models underestimate
largely the binding energy and therefore cannot represent well the
hydrogen-storage property of the crystall which, however,  cannot be
dealt with by the quantum chemistry methods. For many hydrogen molecules adsorbed,
the fragment models even give an opposite trend of the binding energy.

Our calculations for the true crystall structure show that the binding energy
with the metal oxide corner is significantly larger than with the organic linker
due to the metal ion-induced electrostatic force. With many hydrogen molecules
adsorbed the crystall volume and the average binding energy decrease a little bit
because of the small attraction among the hydrogen molecules, which leads to the
regular interlinked H$_2$ nanocage structure for higher H$_2$ density found
previously. For many hydrogen molecules adsorbed on the corner they undergo
a perpendicular-to-parallel transition and prefer to be parallel to the C$_3$
plane.  

The present work shows that the much weaker binding with the organic linker 
can be improved significantly by introducing extra polarization and the resulting
electrostatic interaction: The substitution of F atoms for the H atoms 
can increase the binding energy from 5.50 kJ/mol to 10.39 kJ/mol
which is now very close to the binding energy with the corner. 
This casts a light on possible ways for improving the hydrogen loading of MOFs
for hydrogen storage.

\begin{acknowledgments}
This work was supported by the MOST 973 Project under Grant No. 2011CB922204
and by the National Natural Science Foundation of China under Grant No. 11174220
as well as by the Shanghai Pujiang Program under Grant No. 10PJ1410000.
\end{acknowledgments}


\end{document}